\documentclass[a4paper,11pt]{article}
\pdfoutput=1 

\usepackage{jheppub} 

\usepackage[T1]{fontenc} 

\usepackage{amsmath, amsfonts, amsthm, amssymb, graphicx}

\allowdisplaybreaks[3]

\def\be{\begin{equation}}
\def\ee{\end{equation}}
\def\ba{\begin{eqnarray}}
\def\ea{\end{eqnarray}}
\newcommand{\nn}{\nonumber\\}
\newcommand{\ud}{\mathrm{d}}
\def \pd {\partial}
\def \bfx {{\bf x}}
\def \bfk {{\bf k}}

\title{Gravitational Waves from Oscillon Preheating}

\author[a,b]{Shuang-Yong Zhou,}
\author[a]{Edmund J.~Copeland,}
\author[c]{Richard Easther,}
\author[d]{Hal Finkel,}
\author[a]{Zong-Gang Mou,}
\author[a]{and Paul M.~Saffin}

\affiliation[a]{School of Physics and Astronomy, University of Nottingham, Nottingham, NG7 2RD, UK}
\affiliation[b]{SISSA, Via Bonomea 265, 34136, Trieste, Italy and INFN, Sezione di Trieste, Italy}
\affiliation[c]{Department of Physics, University of Auckland, Private Bag 92019, Auckland, New Zealand}
\affiliation[d]{Argonne Leadership Computing Facility, Argonne National Laboratory, Argonne, IL 60439, USA}

\abstract{Oscillons are long-lived, localized excitations of nonlinear scalar fields which may be copiously produced during preheating after inflation, leading to a possible oscillon-dominated phase in the early Universe. For example, this can happen after axion monodromy inflation, on which we run our simulations. We investigate the stochastic gravitational wave background associated with an oscillon-dominated phase. An isolated oscillon is spherically symmetric and does not radiate gravitational waves, and we show that the flux of gravitational radiation generated between oscillons is also small. However,  a  significant stochastic gravitational wave background may be generated during preheating itself (i.e, when oscillons are forming), and in this case the characteristic size of the oscillons is imprinted on the  gravitational wave power spectrum, which has  multiple, distinct peaks. }


\begin{document} 
\maketitle
\flushbottom

\section{Introduction}

After inflation ends, the energy of the coherent homogeneous inflaton field must be released to reheat the Universe, giving way to the hot Big Bang. This may occur via preheating \cite{Traschen:1990sw, Dolgov:1989us, Shtanov:1994ce, Khlebnikov:1996mc, Kofman:1997yn, Felder:2000hj} where particles  are explosively produced with  non-zero momenta, followed by thermalization. In some cases, high-momentum quanta of the inflaton field can be copiously produced before other particles \cite{Amin:2011hj, Amin:2010dc, McDonald:2001iv, Broadhead:2005hn}. The particle production is seeded by initially minute quantum field fluctuations which are exponentially enhanced via parametric resonance, driven by the homogeneous, oscillating inflaton field. This process is  nonlinear and nonperturbative. Fortunately, as the particle number increases exponentially the occupation numbers of higher momentum states grow rapidly, and the quantum fields can then be well-described by stochastic classical fields \cite{Khlebnikov:1996mc}.

Many interesting non-perturbative field theory phenomena may take place during preheating, including formation of topological or non-topological solitons \cite{Tkachev:1998dc, Kasuya:2000wx, Kusenko:2009cv} as well as quasi-solitons such as oscillons \cite{Bogolyubsky:1976yu, Gleiser:1993pt, Copeland:1995fq, Dymnikova:2000dy, Honda:2001xg, Copeland:2002ku, Farhi:2005rz, Gleiser:2006te, Graham:2006xs, Hindmarsh:2006ur, Fodor:2006zs, Saffin:2006yk, Graham:2006vy, Farhi:2007wj, Hindmarsh:2007jb, Fodor:2008es, Gleiser:2008ty,  Gleiser:2007te, Hertzberg:2010yz, Salmi:2012ta, Amin:2010jq, Amin:2010dc, Amin:2011hj, Amin:2010xe, Amin:2013ika}. Oscillons are spatially localized, temporally oscillating configurations of a nonlinear scalar field theory. First described by Bogolubsky and Makhankov \cite{Bogolyubsky:1976yu}, advances in numerical methods have allowed  these fascinating objects to be studied in great detail  \cite{Gleiser:1993pt, Copeland:1995fq, Honda:2001xg, Copeland:2002ku, Farhi:2005rz, Gleiser:2006te, Graham:2006xs, Hindmarsh:2006ur, Fodor:2006zs, Saffin:2006yk, Graham:2006vy, Hindmarsh:2007jb, Farhi:2007wj, Gleiser:2007te, Salmi:2012ta, Amin:2010jq, Amin:2010xe, Amin:2013ika}. They are not strictly stable, even in the classical limit, as they are not associated with conserved charges (topological or non-topological), but are long-lived due to a trade off in the oscillon potential between the dissipative quadratic terms and an attractive force generated by  the nonlinear terms.

A number of models have been described in which massive oscillons are generated during preheating \cite{Farhi:2007wj, Amin:2011hj, Amin:2010dc, McDonald:2001iv, Broadhead:2005hn, Gleiser:2011xj}.   In particular, Ref.~\cite{Amin:2011hj} showed that in a large class of well-motivated inflationary scenarios oscillons may account for the majority of the energy density in the post-inflationary Universe. This includes the axion monodromy model \cite{Silverstein:2008sg, McAllister:2008hb}. In these scenarios, the Universe undergoes a transient (effective) matter-dominated  phase. For a given inflationary model, a transient  matter-dominated phase  modifies the effective spectral index of the inflationary perturbation spectrum  relative to that produced via immediate thermalization \cite{Mortonson:2010er,Adshead:2010mc}, although this effect degenerates with  other mechanisms that cause the expansion history of the early Universe to depart from a strict radiation-dominated scenario.   

In this paper, we examine possible stochastic  gravitational wave signals generated during this oscillon preheating scenario, to see whether there are any characteristic features  that are -- even in principle -- observable today. Gravitational waves, once generated,   propagate freely even in the very hot, dense early Universe, and are already known to be generated during resonance \cite{Khlebnikov:1997di, Easther:2006gt, Easther:2006vd, GarciaBellido:2007dg,Easther:2007vj,GarciaBellido:2007af, Dufaux:2007pt, Price:2008hq, Dufaux:2008dn, Kusenko:2008zm, Mazumdar:2008up}.   Typically, gravitational waves generated by  preheating after ``high scale'' inflation are found in the MHz to GHz range, well beyond the scope of current or planned interferometric detectors. However, their importance as a potential probe of the post-inflationary Universe and inflation models has motivated the study of the possible stochastic backgrounds generated during preheating and provided impetus for the development of novel detector technologies.  

In what follows, we will show that the characteristic scale of the oscillons is imprinted on the gravitational wave background associated with the resonant phase that generates them, giving rise to power spectra with organized broad peaks. Conversely, once initial transients decay, oscillons are rather spherically symmetric, so an isolated oscillon does not radiate gravitational waves. We also compute the gravitational wave emission generated between multiple oscillons, and show that this is very small, so an oscillon-dominated phase itself does not generate significant gravitational waves. However, the amount of gravitational wave energy produced when the oscillons are forming is typical of that of a usual preheating scenario.

The outline of the paper is as follows. In Section \ref{sec:ospre}, we briefly review oscillon dynamics  and introduce the oscillon preheating scenario. In Section \ref{sec:gwana}, we analytically estimate the gravitational wave production from the oscillon-dominated phase and show that the production is highly suppressed when oscillons are properly formed. In Section \ref{sec:gwnu} we describe our numerical algorithm and present the results of our lattice simulations, and we conclude in Section \ref{sec:conclu}. 

\section{Oscillons and Oscillon Preheating} \label{sec:ospre}

Oscillons can form in a variety of models (see e.g.~\cite{Amin:2011hj, Farhi:2005rz, Gleiser:2007te, Gleiser:2011xj}), but given that an oscillon-dominated Universe can follow inflation driven by a single minimally-coupled scalar field, we restrict our attention to this scenario \cite{Amin:2011hj}.   Oscillons can form when the scalar potential  ``opens up''~\cite{Amin:2010jq}. Suppose the potential has the form
\be
V(\phi) = \frac{m^2}{2} \phi^2 + V_{\rm nl}(\phi)  ,
\ee
with a minimum at $V(\phi=0)=0$\;\footnote{A minimum at non-zero $\phi$ may be shifted to the origin by a field redefinition.} and is symmetric under $\phi\to -\phi$, then the `opening up' condition simply means~\cite{Amin:2010jq}
\be \label{Vnlcondition}
V_{\rm nl}(\phi) < 0 ~~~~{\rm for~some~region~of~}\phi .  
\ee
To understand this condition, we will need an approximate analytic description of oscillon physics.  For oscillons much smaller than the characteristic spacetime curvature, the Klein-Gordon equation is simply
\be \label{kgeq}
\ddot{\phi} - \nabla^2 \phi + m^2 \phi + V'_{\rm nl}(\phi) = 0 .
\ee
An oscillon usually has a dominant oscillating frequency which we write as
\be \label{firstosc}
\phi(x)=\Phi(\bfx)\cos(\omega t) ,
\ee
where $\Phi(\bfx)$ is a positive localized function that falls off quickly far from the center.  With this ansatz, far from the center of the oscillon the field equation reduces to $(m^2-\omega^2) \Phi  \simeq \nabla^2 \Phi$. Since $\nabla^2 \Phi$ has to be small but positive in this region, the angular frequency satisfies
\be \label{olm1}
\omega \simeq  m ~~~ {\rm and} ~~~\omega < m   ,
\ee
showing that the mass parameter sets the dominant oscillation frequency.  In the central region, the nonlinear term $V'_{\rm nl}$ is important and $\nabla^2 \Phi$ is negative. In combination with Eq.~(\ref{olm1}) this leads to $-\omega^2 \Phi - \nabla^2 \Phi + m^2\Phi > 0$.  Since Eq.~(\ref{kgeq}) should be satisfied for any $t$,  we get that $V'_{\rm nl}(\Phi)= -(-\omega^2 \Phi - \nabla^2 \Phi + m^2\Phi)$ is negative in the oscillon center if $\cos(mt)= 1$, and $V'_{\rm nl}(-\Phi)$ is positive it $\cos(mt)= -1$. Integrating over the field from $0$ to $\phi$, we then obtain condition (\ref{Vnlcondition}). Physically, if the interaction term  is shallower than the mass term (e.g.~$-\phi^4$), it induces an attractive force between ``particles'', which can balance the dissipative quadratic terms, and rendering oscillon formation energetically preferred.

Our ansatz (\ref{firstosc}) captures the dominant oscillating mode of the oscillon. However, an oscillon also has higher order harmonics; e.g. $\cos(2m t)$, $\cos(3mt)$, $\cdots$ \cite{Fodor:2008es, Salmi:2012ta}. We can see this nicely in the small amplitude limit \cite{Fodor:2008es} where the smooth configuration scale of the oscillon is much larger than its natural oscillation period, i.e.,
\be
\frac{\pd}{\pd t} \sim m \sim  \frac{\pd}{\epsilon\pd x^i}~~~ {\rm and}~~~\phi\sim\epsilon m   ,
\ee
$\epsilon$ being a small parameter. To count orders, we expand the field as
\be
\phi(x) = \sum_{n=1}^{\infty} \epsilon^n \phi_n(x) ,
\ee
and redefine the spatial coordinates $\bar{x}^i = \epsilon x^i$.
Assuming $V_{\rm nl}(\phi)=\sum_{a>2} g_a\phi^a/a$ ($g_a$ being coupling constants), the equations of motion for the modes $\phi_n$ are
\be
\ddot{\phi}_n+ m^2 \phi_n - \nabla_{\bar{x}}^2\phi_{n-2} + \sum_{a>2}~\sum_{n_1+n_2...+n_{a-1}=n}g_a\phi_{n_1}\phi_{n_2}...\phi_{n_{a-1}} = 0
\ee
where $n$ goes from 1 to $\infty$ and we  define $\phi_{-1}=\phi_{0}=0$. We can solve this set of equations perturbatively, and the first two equations are
\begin{align}
\label{phi1eom}
\ddot{\phi}_1+ m^2 \phi_1  & = 0  ,
\\
\label{phi2eom}
\ddot{\phi}_2+ m^2 \phi_2 +g_3\phi_1^2 & = 0    .
\end{align}
Eq.~(\ref{phi1eom})  gives $\phi_1=\Phi_1(\bar{\bf x})\cos(mt)$, recovering our ansatz (\ref{firstosc}). (To be fully rigorous here, we should have allowed for a spatially dependent phase in the cosine, but the boundedness of the whole solution forces it to be constant \cite{Fodor:2008es}, and we have set it to zero.) Inserting our solution for  $\phi_1$,  Eq.~(\ref{phi2eom}) reduces to a forced harmonic oscillator, with a forcing angular frequency $2m$, with the solution
\be\label{phi2result}
\phi_2=\Phi_{21}(\bar{\bf x})\cos(mt)+\Phi_{22}(\bar{\bf x})\sin(mt) +\frac{g_3 \Phi_1(\bar{\bf x})^2}{6m^2} [\cos(2mt) - 3]  .
\ee
By induction we see that the higher order equations are all forced oscillators, and at each step in this hierarchy  the (largest) forcing angular frequency increases by $m$, which in turn increases the (largest) angular frequency for $\phi_n$ by $m$.  If the potential is symmetric under $\phi\to -\phi$ around the minimum, the odd order coupling constants vanish ($g_{2a+1}=0$). From Eq.~(\ref{phi2result}), we see that then the second harmonic $\cos(2mt)$ vanishes and this can be generalized to higher orders, so higher harmonics with the angular frequency $2n m$ all vanish. The existence and properties of these various harmonics away from the small amplitude limit have been investigated numerically in \cite{Salmi:2012ta}, with results consistent with this small amplitude analysis. These results concern an oscillon's behavior in the time direction. As $m$ is the ``dominating'' mass scale in this Lorentz invariant model, a similar $m$ modulated behavior may be expected for the spatial directions.

From an inflationary perspective, the ``opening up'' condition (\ref{Vnlcondition}) is relatively easy to satisfy as current data  disfavor the quadratic potential relative to flatter potentials \cite{Komatsu:2010fb, Mortonson:2010er, Martin:2010kz, Ade:2013rta}.    For a consistent description of preheating with a single scalar, we must also assume the inflaton's self-couplings, at least initially, are more important than its couplings to other fields.    Inspired by a number of string and supergravity inflation models \cite{Silverstein:2008sg, McAllister:2008hb, Dong:2010in, Kallosh:2010xz, Kallosh:2010ug}, a family of potentials have been proposed that yields  an oscillon preheating scenario whilst satisfying current observational constraints \cite{Amin:2011hj}
\be \label{modelpot}
V(\phi) = \frac{m^2M^2}{2\alpha} \left[ \left( 1+\frac{\phi^2}{M^2}\right)^\alpha -1\right] ,
\ee
where $0\leq \alpha\lesssim 0.9$ and $M\lesssim 0.05M_P$ ($M_P^2=1/8\pi G$). In particular, $\alpha = 1/2$ is the axion monodromy model \cite{Silverstein:2008sg, McAllister:2008hb}. When $\phi/M$ is large, the potential can be approximated by $V(\phi)\simeq m^2 M^{2-2\alpha} \phi^{2\alpha}/2\alpha$. On the other hand, when $\phi/M$ is small, we have $V(\phi)\simeq m^2\phi^2/2$. For this potential very massive oscillons can be copiously produced after inflation, giving rise to an oscillon-dominated epoch before thermalization. In this paper we ask whether oscillon preheating leaves any ``footprints'' in today's Universe, particularly in the gravitational wave background it generates.

\section{Gravitational Waves: Analytical study} \label{sec:gwana}

There are three possible ways where gravitational waves can arise as a consequence of an oscillon-dominated phase in the early Universe -- during their formation, during the oscillon dominated phase itself, and during their  decay and the subsequent thermalization of the Universe. We cannot address the last issue, since this process is beyond the scope of the model we are analyzing. The first process will be driven by strong nonlinear, nonperturbative dynamics and will be investigated numerically in the following section.  In this Section we give a semi-analytical treatment of gravitational wave production in the oscillon-dominated phase itself.

\begin{figure}
\begin{center}
\includegraphics[height=5in,width=5in]{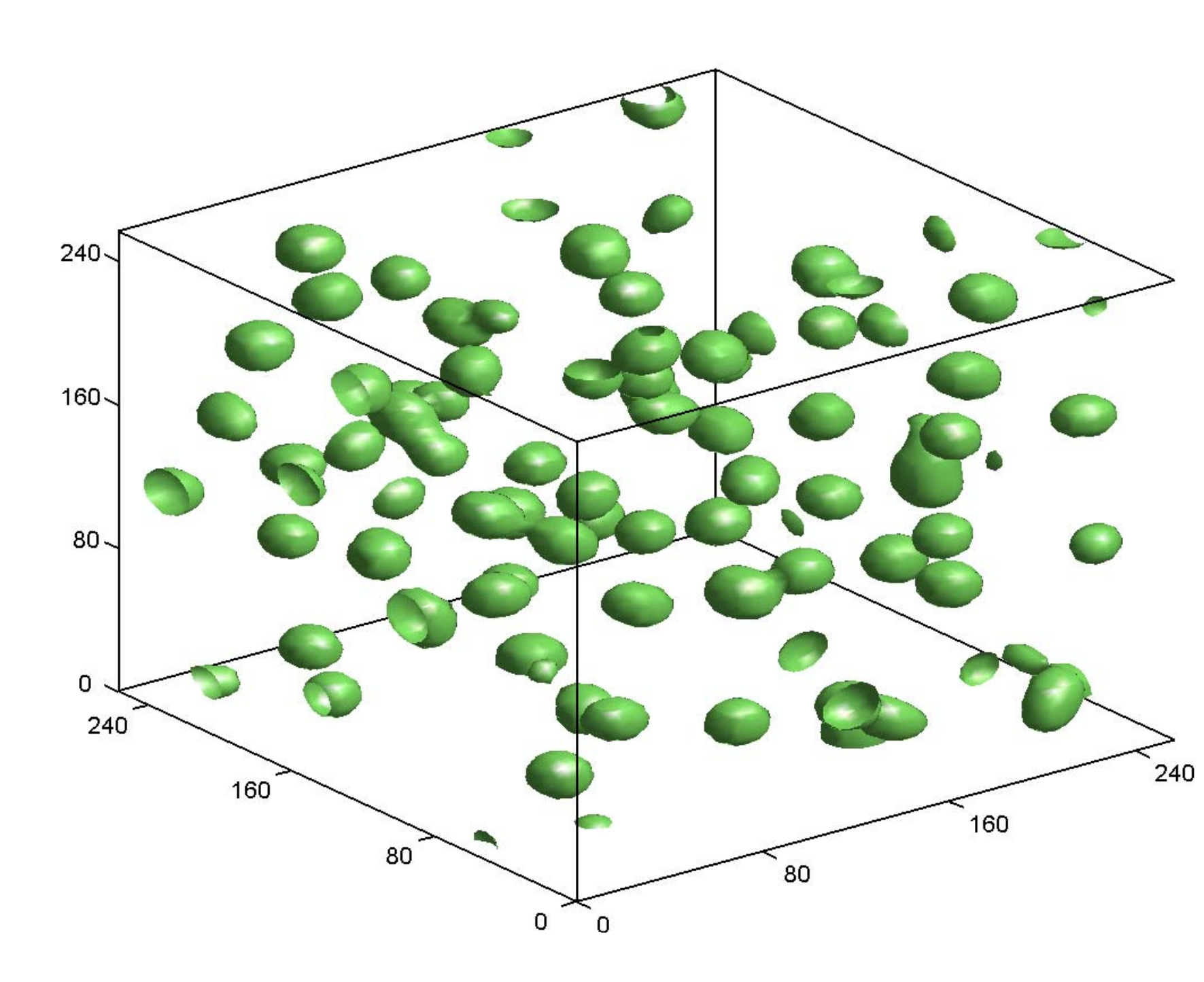}
\caption{A late time snap shot of the energy density in oscillon preheating. The model is that of Eq.~(\ref{modelpot}) with $\alpha=1/2$ and $M=0.01M_P$. The box size is $L=50/m$ and the energy density isosurface is taken at a value 5 times the average energy density. } \label{fig:osc3d}
\end{center}
\end{figure}

In simulations of oscillon formation, the resulting oscillons are essentially at rest with respect to the background spacetime.  This makes sense: we can view an oscillon as a collection of a scalar particles (i.e., a system with a large number of degrees of freedom) which are generated from a homogeneous field configuration with a vanishing total momentum, so the overall momentum of the generated oscillon should be highly suppressed.  Consequently, as a good approximate description, we can  assume that oscillons are not moving. Secondly, oscillons are formed by causal interactions, well within the Hubble horizon, so we estimate the production of gravitational waves from oscillons in Minkowski space, where we can use standard gravitational wave formulae, summarized in Appendix~\ref{sec:formgw}. Thirdly, gradient energy is minimized by spherically symmetric profiles, and numerical simulations show that the oscillons approach a symmetric state as they evolve  (see Figure (\ref{fig:osc3d})).  Consequently we assume spherically symmetric oscillons. (In Appendix~\ref{sec:nssl}, we show that our results also apply for the case of small deviations from perfect spherical symmetry.) We will see that the property of spherical symmetry has important implications for the gravitational wave production in the oscillon-dominated phase. Fourthly, from numerical simulations we see that oscillons are well separated in distance. This is not surprising since oscillons originate from a homogenous surrounding to become highly concentrated energy lumps, leaving large low energy density voids in between. Finally, the nonlinear potential makes it difficult to analytically describe the spatial profile of the oscillon and  we will assume a Gaussian spatial profile for convenience.

Consider a Minkowski space homogeneously distributed with the idealized oscillons described above.  According to Birkhoff's theorem, there is no gravitational wave emission from a spherically symmetric source, oscillating or not. As a consistency check, one can substitute the ansatz $\phi(t,{\bf x}) \propto e^{ -m_o^2 {\bf x}^2/2 }\cos(\mu t)$ (where $m_o$ and $\mu$  characterize respectively the width and angular frequency of one harmonic of the oscillon) into Eq.~(\ref{formula3}) and find that the emitted power in gravitational waves indeed vanishes.

Next, we want to consider the gravitational wave production from superposition of multiple oscillons. Due to the spherical symmetry of the oscillons, we expect this contribution to be suppressed when oscillons are well separated. It is, however, unclear how efficient the suppression is in terms of the gravitational wave emission. When oscillons are well-separated, we may superpose their field configurations  and consider the following ansatz
\be
\phi(t,\bfx) = \sum_{a=1}^\infty  A_a e^{-\frac12 m_o^2(\bfx -{\bf d}_a)^2} \cos(\mu t+\varphi_a)  ,
\ee
where $A_a$ is the amplitude of the center of the oscillon, ${\bf d}_a$ is the spatial position of the oscillon and $\varphi_a$ the initial phase.  The resulting gravitational waves have angular frequency $2\mu$, since  schematically the perturbative Einstein equation is $\Box h_{ij} \sim T_{ij} \sim \pd_i \phi \pd_j \phi \sim \cos^2(\mu t+\varphi_a)$, where $h_{ij}$ is the perturbative metric. Substituting this ansatz into the formula (\ref{formula3}), after some tedious Fourier transforms and algebra we get the emitted power
\be \label{gwpess}
\frac{\ud P_{2\mu}}{\ud \Omega} =  \frac{\ud P_2}{\ud \Omega} + \frac{\ud P_4}{\ud \Omega}
\ee
with
\begin{align}
\frac{\ud P_2}{\ud \Omega} &= \frac{\pi^2 G}{32} \sum_{a>b} (A_a A_b)^2 (\mu{\bf d}_{ab})^2(m_o{\bf d}_{ab})^2\sin^4(\hat{\bfx},{\bf d}_{ab}) e^{-\frac12 m_o^2 {\bf d}_{ab}^2 -\frac{\mu^2}{m_o^2}}  , \\
\frac{\ud P_4}{\ud \Omega}&=   \frac{\pi^2 G}{16}\sum_{(a>b)\neq(c>d)}  A_a A_b A_c A_d (\mu{\bf d}_{ab})^2 (m_o{\bf d}_{cd})^2 S_{abcd}(\hat{\bfx},{\bf d},\varphi) e^{-\frac14 m_o^2 ({\bf d}_{ab}^2+{\bf d}_{cd}^2) - \frac{\mu^2}{m_o^2}}   ,
\end{align}
where $\sum_{a>b}$  implies summation over $a$ and $b$ that satisfy $a>b$ and $\sum_{(a>b)\neq(c>d)}$ is over $a$, $b$, $c$ and $d$ that satisfy $a>b$, $c>d$ and $(a,b)\neq (c,d)$. We have  introduced $\hat{\bfx}=\bfx/|\bfx|$, ${\bf d}_{ab}={\bf d}_{a}-{\bf d}_{b}$, $\varphi_{ac}=\varphi_a-\varphi_c$, and
\begin{align}
S_{abcd}(\hat{\bfx},{\bf d},\varphi) \equiv \Big[ &\left[\cos({\bf d}_{ab},{\bf d}_{cd})-\cos(\hat{\bfx},{\bf d}_{ab})\cos(\hat{\bfx},{\bf d}_{cd})\right]^2 - \frac12\sin^2(\hat{\bfx},{\bf d}_{ab}) \sin^2(\hat{\bfx},{\bf d}_{cd})  \Big] \nn
 &~~\cdot\cos( \frac{1}{\sqrt{2}}\hat{\bfx}\cdot (\mu{\bf d}_{ac}+\mu{\bf d}_{bd})+\varphi_{ac}+\varphi_{bd}) .
\end{align}
where  $\sin(\hat{\bfx},{\bf d}_{ab})$, denotes the sine of the angle between $\hat{\bfx}$ and ${\bf d}_{ab}$. $\ud P_2/\ud \Omega$ is the contribution from pair-wise oscillons and $\ud P_4/\ud \Omega$ comes from the four-oscillon contribution, and because of the structure $T_{ij} \sim \pd_i \phi \pd_j \phi$, there is no other contributions. 

The dominant frequency of oscillons is slightly smaller than $m$ (the mass parameter in the potential) and most of the energy-density is localized within a radius  $r \simeq m^{-1}$ of the oscillon core, so as a good estimate we can set $m_o=\mu=m$ in the rest of this Section. (The results are similar if we consider the higher harmonics $m_0=\mu=2m, 3m,...$.)

We can see that either the two-oscillon or the four-oscillon contribution is exponentially suppressed (i.e. $m_o^2 {\bf d}_{ab}^2 \gg1$), since in the oscillon preheating scenario the separation of oscillons are usually $\gtrsim m^{-1}$. For the two-oscillon contribution, this simply means that the leading contribution is from immediately neighboring oscillon pairs. Thus the power density of gravitational wave emission (the power of emission (\ref{gwpess}) divided by the space volume) from $\ud P_2/\ud \Omega$ is highly suppressed.

The suppression of the four-oscillon contribution is less straightforward to see, even if we  intuitively expect it to be small. The exponential suppression above is by $(m{\bf d}_{ab})^2$ and $(m{\bf d}_{cd})^2$. While these suppressions restrict attention to small $(m{\bf d}_{ab})^2$ and $(m{\bf d}_{cd})^2$ contributions, they do not suppress large separations between oscillon $a$ and oscillon $c$ and between  oscillon $b$ and oscillon $d$. That is, in the summation $\sum_{(a>b)\neq(c>d)}$, it might appear that we  have to consider a large number of four-oscillon contributions (the ones with small $(m{\bf d}_{ab})^2$ and $(m{\bf d}_{cd})^2$ but (arbitrary) large $(m{\bf d}_{ac})^2$ and $(m{\bf d}_{bd})^2)$. If this were the case, the four-oscillon contribution to the power density of gravitational wave emission would diverge,  contradicting the usual principle that energy is extensive and can be considered as a sum of those of the sub-volumes. To see how locality is respected, focus on the factor $\cos( \hat{\bfx}\cdot (m{\bf d}_{ac} + m{\bf d}_{bd}) / \sqrt{2}+\varphi_{ac}+\varphi_{bd})$, which is the only place where ${\bf d}_{ac}$ and ${\bf d}_{bd}$ enter the formula. When $(m{\bf d}_{ab})^2$ and $(m{\bf d}_{cd})^2$ are small and ${\bf d}_{ac}$ and ${\bf d}_{bd}$ are large, ${\bf d}_{ac}$ and ${\bf d}_{bd}$ have to be almost parallel, in which case
\be
\cos( \frac{1}{\sqrt{2}} \hat{\bfx}\cdot (m{\bf d}_{ac}+m{\bf d}_{bd})+\varphi_{ac}+\varphi_{bd})\backsimeq\cos(\sqrt{2}|m{\bf d}_{ac}|\cos\theta+\varphi_{ac}+\varphi_{bd}) ,
\ee
$\theta$ being the angle between $\hat{\bfx}$ and ${\bf d}_{ac}$.  When $|m{\bf d}_{ac}|$ is large this cosine factor oscillates rapidly about zero when $\cos\theta$ varies (Note that $\ud \Omega = \ud \cos\theta \ud \phi$.). Physically, the energy emission is obtained by integrating a finite solid angle. For a given solid angle, large enough $|m{\bf d}_{ac}|$   this cosine oscillation suppresses any net contribution.  Thus only configurations with  four adjacent oscillons can contribute and these, as mentioned above, are highly suppressed. 

Therefore we conclude that gravitational wave production is negligible during the oscillon dominated phase itself, as the oscillons are rather spherically symmetric when properly formed and well separated in distance.

\section{Gravitational Waves: Numerical simulations} \label{sec:gwnu}

In this Section, we compute the gravitational wave production during the resonance and oscillon-formation phase.  Our numerical code is based on PSpectRe~\cite{Easther:2010qz}, a pseudo-spectral code for simulating the evolution of coupled scalar fields on an expanding background, upgraded with the capacity of computing gravitational wave power spectra. The initial inhomogeneous seeds of the scalar field are generated similar to DEFROST~\cite{Frolov:2008hy}. They are originated from the Gaussian, vacuum quantum fluctuations of the inflaton field after the end of inflation, and become (semi-)classical because the rapid parametric resonance in preheating leads to large occupation numbers for the relevant modes (for more details, see~\cite{Easther:2010qz} for the algorithm of their generation). The scalar field evolution is computed using a second-order-in-time symplectic integration scheme with a fixed time step. The transverse-traceless part of the first order perturbative metric is computed using the method described in~\cite{Easther:2007vj} -- with the Fourier modes of the metric perturbations (and their sources in $T_{\mu\nu}$) evolved in Fourier space, via a fourth-order-in-time Runge-Kutta scheme.  To first order in the metric perturbations, the entire combined system is self-consistent: the background evolution is self-consistently computed from the scalar field and the evolution equations used to evolve the metric perturbations are valid at all scales and do not assume any particular expansion history.

In the previous section, we have shown that when oscillons are properly formed there is no significant gravitational wave production. However, before the oscillon-dominated phase, there is a phase of parametric resonance followed by a short phase of nonlinear ``re-scattering'', when the homogeneous inflaton modes fragment into higher momentum modes. These phases, especially the ``re-scattering'' phase, are quite violent, so we can expect a significant stochastic gravitational wave background to be generated, much like a typical preheating scenario. As we will see, this is indeed the case and there are some novel features tied to the forming of oscillons.

The results  presented here are for the model of Eq.~(\ref{modelpot}) with $\alpha=1/2$ (i.e., the axion monodromy model) and $M=0.01M_P$. The oscillon field mass $m$ is determined by matching to the curvature power spectrum extracted from the temperature anisotropies in the cosmic microwave background observed today~\cite{Komatsu:2010fb}\footnote{We used the amplitude of the power spectrum obtained from the WMAP 7-year analysis  \cite{Komatsu:2010fb}, but our results are not strongly dependent on this parameter.}
\be \label{cmbdr}
\Delta^2_R = \left.\frac{H_*^2}{(2\pi)^2}\frac{H_*^2}{\dot{\phi}_*^2}\right|_{\rm CMB} = \frac{1}{96\pi^2\alpha^3} \left(\frac{m}{M_P} \right)^2 \left(\frac{M}{M_P}\right)^{2-2\alpha} (4\alpha \mathcal{N})^{1+\alpha}= 2.4\times 10^{-9}
\ee
where $\mathcal{N}$, the number of e-folds before the end of inflation, is set to  55 here. This gives $m=3.1\times 10^{-5}M_P$ and the inflation scale can be estimated by setting $\phi=M$, which is about $10^{15} {\rm GeV}$. We have checked the stability of our results against changes in lattice and box sizes as well as physical parameters such as $M$. The results  presented here are derived from simulations on a $256^3$ lattice with box size $L=50/m$ (the initial Hubble horizon being about $270/m$), unless otherwise stated.

\begin{figure}
\begin{center}
\includegraphics[height=4.5in,width=6in]{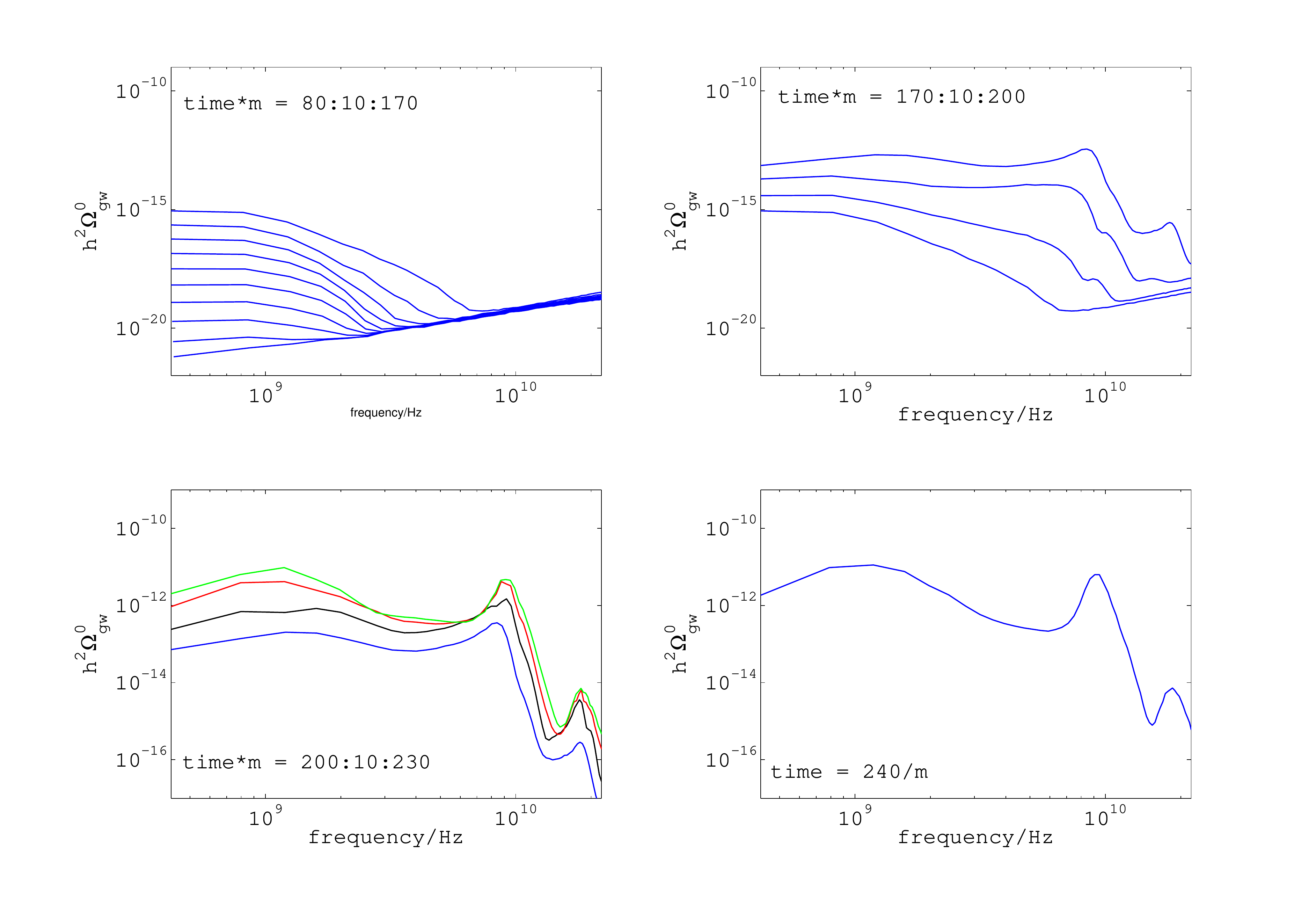}
\caption{Evolution of the gravitational wave power spectrum.  The lattice is $256^3$, the box size is $L=50/m$ and the evolution time shown is from $t=80/m$ to $t=240/m$. There are (at least) four distinct stages: (top-left) - steady ``pumping'' (linear parametric resonance) in low momentum modes; (top-right) - rapid growth (nonlinear ``re-scattering''), oscillons forming; (bottom-left) - oscillons stabilizing, peaks and troughs becoming significant; (bottom-right) - long stable stage, very slow growth.} \label{fig:fourstages}
\end{center}
\end{figure}

\begin{figure}
\begin{center}
\includegraphics[height=3.6in,width=5.5in]{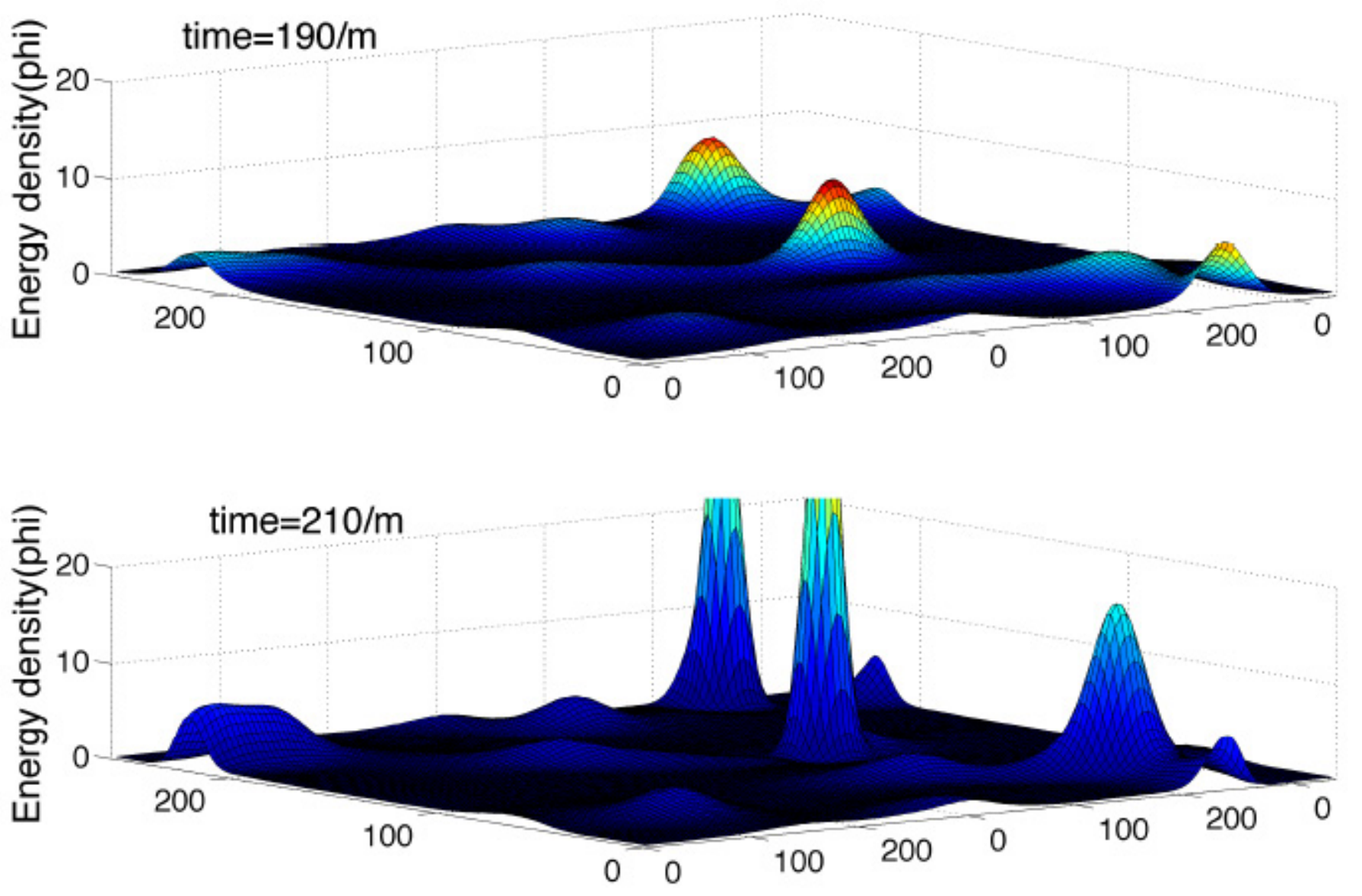}
\caption{The energy densities of the same slice of the box at two different times. The model is that of Eq.~(\ref{modelpot}) with $\alpha=1/2$ and $M=0.01M_P$. The lattice is $256^3$ and the box size is $L=50/m$. The top is taken at the late time of the rapid growth stage and the bottom is taken at the early time of the stabilizing stage. We can see that at the onset of the stabilizing stage oscillons become very massive, dominating the energy density of the Universe.} \label{fig:twoslices}
\end{center}
\end{figure}

Figure (\ref{fig:fourstages}) shows the time evolution of the gravitational wave power spectrum, plotted using the present-day frequency and fractional energy density in gravitational waves, assuming instantaneous thermalization at the instant the spectrum is computed (see e.g.~\cite{Dufaux:2007pt} for the conversion.)   If there is a long oscillon-dominated phase, this last assumption will need to be revisited, as we discuss below.   The power spectrum curve grows upwards with time, which is labelled in units of $1/m$. In what follows, we describe in succession  four stages that can be easily identified.  A significant production of gravitational waves begins at around $t=80/m$, with an initial steady growth, characterized by the ``pumping'' of the low frequency modes (top-left graph). This is a stage of linear parametric resonance. Here fluctuations of the inflaton field are small compared to the zero-mode oscillation.  During this stage the total power fluctuates as newly created gravitational waves can constructively and destructively interfere with those that are already there, although the overall trend is toward greater power.  The second stage is rapid growth  (top-right), as  oscillons  emerge and the zero-mode oscillation of the inflaton decays rapidly. This short stage is sometimes called nonlinear ``re-scattering'' for preheating scenarios when oscillons are not formed. Interesting peaks and troughs starts to emerge in the power spectrum at the later time of this stage. The rapid growth stage is followed by a stabilization, as oscillons further drain energy from the environment, with existing oscillons becoming more massive and new smaller ones emerging  (bottom-left graph).  Figure (\ref{fig:twoslices}) shows the energy density as a function of position on a spatial slice at two close times, one taken at the late time of the rapid growth stage and the other taken at the early time of the stabilizing stage. We can see that oscillons in the stabilizing stage are much more massive than in the rapid growth stage. More interestingly, back to Figure (\ref{fig:fourstages}), the structure of peaks and troughs becomes more evident at the stabilizing stage.  Finally,  we have a long stable stage, when  oscillon formation is complete and the gravitational wave power spectrum grows very slowly (bottom-right graph).

\begin{figure}
\begin{center}
\includegraphics[height=3.43in,width=5in]{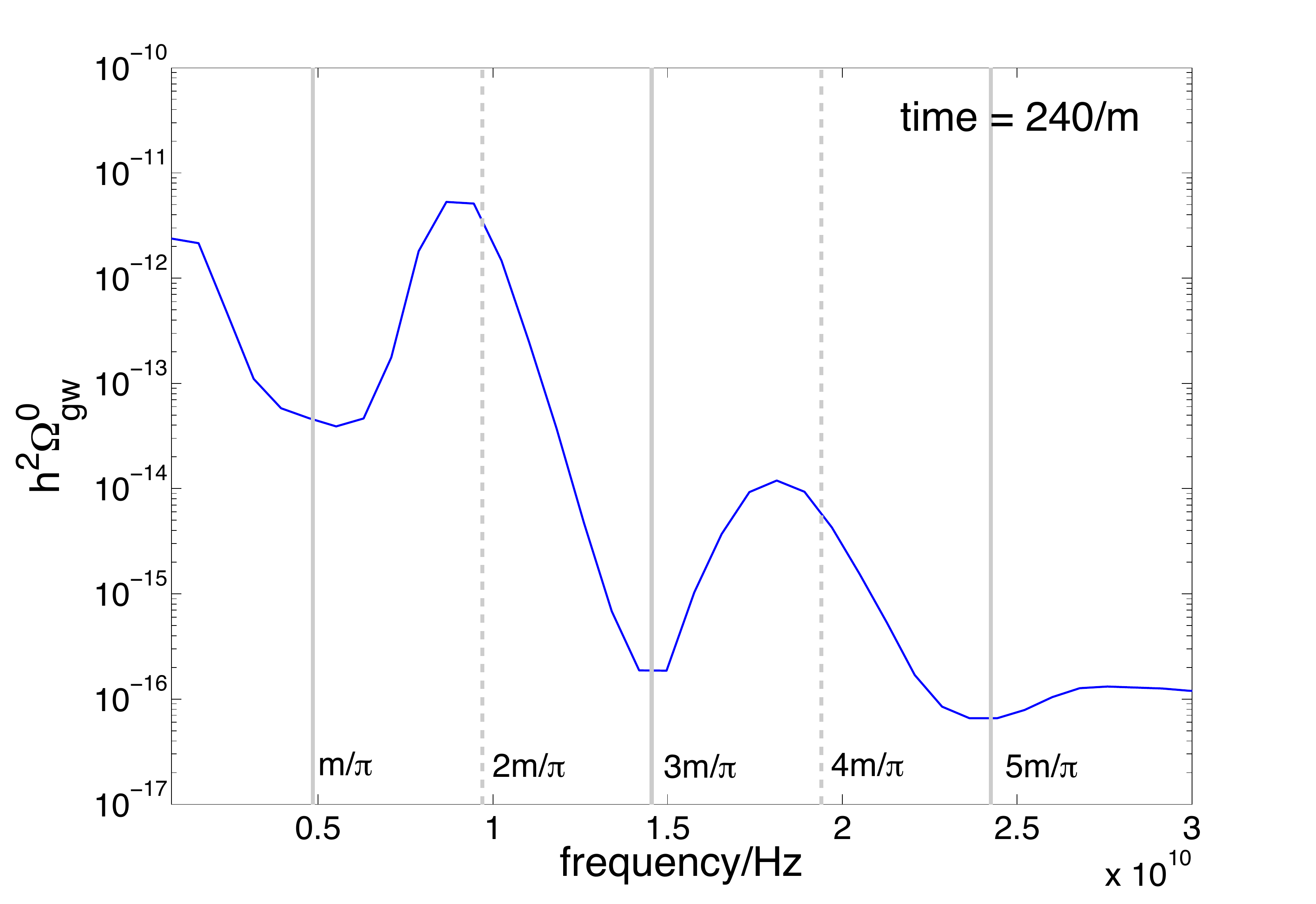}
\caption{Peaks and troughs of the gravitational wave power spectrum. The model is that of Eq.~(\ref{modelpot}) with $\alpha=1/2$ (the axion monodromy model) and $M=0.01M_P$. The lattice is $256^3$, plotted at $t=240/m$. The box size is $L=25/m$, instead of $L=50/m$, so as to resolve more high momentum modes. The vertical lines correspond to the gravitational wave frequencies associated with the different harmonics of the oscillon, which are twice the frequencies of the oscillon harmonics. Note that the values indicated for the vertical lines ($m/\pi,3m/\pi,5m/\pi,...$) should be understood as being multiplied by a redshift factor ($\sim 10^{-30}$). Even order oscillon harmonics ($2m/\pi, 4m/\pi,$ ..., in terms of the gravitational wave frequency; the dotted lines) are absent in our model, because of the symmetry $\phi\to -\phi$. The matches of the vertical lines with the troughs indicate that oscillons are suppressing the gravitational wave production at those frequencies.  } \label{fig:highf}
\end{center}
\end{figure}

In order to better investigate the peaks in the spectrum we repeat these calculations in a smaller  box ($25/m$) to resolve higher frequencies, with results shown in Figure (\ref{fig:highf}). As we discussed in Section \ref{sec:ospre}, the oscillon has a dominant angular frequency of  $\omega \sim m$, as well as higher order harmonics with  angular frequencies  $3m, 5m,7m, ...$. The even order harmonics are absent, because the model given by Eq.~(\ref{modelpot}) is symmetric under $\phi \to -\phi$.  Interestingly, we see the same structure in the gravitational wave power spectrum -- the spacing of the troughs has the same pattern as the oscillon harmonics, as also shown in Figure (\ref{fig:highf}). 

We have shown in Section \ref{sec:gwana} that the gravitational wave emission from spherically symmetric oscillons is highly suppressed. Therefore, the fact that these troughs and peaks begin to appear when oscillons are becoming massive and increasingly spherically symmetric (see Figure (\ref{fig:fourstages}) and Figure (\ref{fig:twoslices})) seems to suggest a simple explanation: the troughs are forming because spherically symmetric oscillons suppress the power growth at those bands. Note that, as we have mentioned in Section \ref{sec:gwana}, the angular frequencies of the gravitational waves these different harmonics excite are twice the corresponding oscillon frequencies. In Figure (\ref{fig:highf}), we have also plotted the estimated gravitational wave frequencies (the angular frequencies divided by $2\pi$) associated with the different harmonics of the oscillon (the vertical lines). Note that the values indicated for the vertical lines ($m/\pi,3m/\pi,5m/\pi,...$) should be understood as being multiplied by a redshift factor ($\sim 10^{-30}$), which we have omitted to avoid cluttering.

Oscillons are quasi-stable  but our gravitational wave spectra are plotted using a transfer function which assumes instantaneous thermalization.  During the matter-like oscillon-dominated phase the gravitational waves will redshift relative to the overall density of the Universe, reducing their energy density, but their present-day frequency is also reduced, since the comoving horizon size is growing more slowly.  Since the overall duration of the oscillon dominated phase is unknown (it depends on couplings between the inflaton and other fields) we can only compute this correction parametrically.   If the Universe grows $\mathcal{N}_{osci}$ e-folds larger during the oscillon-dominated phase, gravitational wave  frequencies are redshifted by a factor $e^{\mathcal{N}_{osci}/4}$ while the power spectrum amplitude is diluted by a factor of $e^{\mathcal{N}_{osci}}$.  Also, a modified expansion history affects the evaluation of $\mathcal{N}$ which determines the value of $m$ via Eq.~(\ref{cmbdr}) and thus affects the gravitational wave power spectrum. For example, if $\mathcal{N}_{osci}\sim 7$ and assume the gravitational wave production is not significant in the decay of oscillons, considering all these effects, we would have $f_{\rm gw}^{typical}\sim 10^8$ Hz and $h^2\Omega_{\rm gw}^{typical}\sim 10^{-14}$.

\section{Conclusions}\label{sec:conclu}

We have studied the stochastic gravitational wave background associated with an oscillon-dominated phase in the early Universe. We see significant gravitational wave production as oscillons are being formed, and  the resulting spectrum has a specific and characteristic structure, with multiple peaks and troughs as shown in Figure~(\ref{fig:highf}).  Further, the frequencies associated with the peaks and troughs are matches to those that contribute to the internal dynamics of the oscillon. The significant production of gravitational waves happens in the early phase of oscillon preheating, when parametric resonance and ``re-scattering'' take place and the homogenous inflaton modes fragment into higher momentum modes, i.e., when oscillon preheating resembles a usual preheating scenario. Conversely, when oscillons are properly formed and dominating the Universe, the gravitational wave production is not significant. Both the appearance of multiple troughs in the power spectrum and the lack of gravitational energy emission in the oscillon-dominated phase itself seems to stem from the fact that oscillons, after formation, morphologically evolve towards spherical symmetry. 

Generically, considering stochastic gravitational backgrounds produced at the end of inflation leads us into a somewhat paradoxical position. Intellectually, these are among the most fascinating topics in cosmology, and provide a potential window into the Universe at the end of inflation. From a practical perspective, however, they occur at frequencies far beyond the capabilities of present-day gravitational wave detectors, or even any reasonable extrapolation of interferometer-based technologies.  Consequently, detecting a stochastic gravitational background generated at the end of ``high scale'' inflation will require some new technology, and the signal discussed here is no exception. 

This analysis deepens our understanding of the phenomenology of  monodromy inflation and its generalizations. From a theoretical perspective, monodromy inflation is one of the most solidly-motivated inflationary scenarios \cite{Silverstein:2008sg,McAllister:2008hb} and a good match with current data \cite{Ade:2013uln}. In addition to a possible oscillon-dominated phase, monodromy inflation  is also associated with a possible modulated power spectrum, a possibility which can be usefully constrained with modern astrophysical datasets \cite{Peiris:2013opa} and would be correlated with a substantial 3-point function \cite{Flauger:2009ab}. Consequently, it is worthwhile to explore the properties of   this model. The rich  phenomenology of monodromy inflation also underlines the importance of carefully considering the full consequences of inflationary models, in order to better understand -- and eventually test -- their predictions. 

The appearance of multiple troughs, as we argued, is due to formation of spherically symmetric oscillons and thus presumably is quite universal in models where oscillons can form and dominate the Universe. It would be interesting to see whether oscillon preheating can arise in any ``low scale'' inflation models and thus leads to signals detectable in the near future.

~\\
{\bf Acknowledgment}
We would like to thank Mustafa~A.~Amin, Paul Tognarelli and Wei Xue for helpful discussions. EJC would like to thank the Royal Society, STFC and Leverhulme Trust for financial support.  SYZ  acknowledges partial financial support from the European Research Council under the European Union's Seventh Framework Programme (FP7/2007-2013) / ERC Grant Agreement no 306425 ``Challenging General Relativity", and from the Marie Curie Career Integration Grant LIMITSOFGR-2011-TPS Grant Agreement no 303537. HJF is supported by the U.S. Department of Energy, Basic Energy Sciences, Office of Science, under contract \# DE-AC02-06CH11357. We would like to thank the High Performance Computing facility at the University of Nottingham for running of our simulations.

\appendix

\section{Formulae of the gravitational wave power spectrum} \label{sec:formgw}

Given a source with energy momentum tensor $T_{\mu\nu}(t,\bfx)$, the gravitational energy emitted per solid angle is given by \cite{weinbergbook}
\be \label{formula1}
\frac{\ud E}{\ud \Omega} =  \frac{G}{2\pi^2} \Lambda_{ij,lm}(\hat{k}) \int_0^{\infty}\omega^2 \ud \omega T^{ij}(\omega,\bfk) (T^{lm}(\omega,\bfk))^*  .
\ee
where $\bfk =  \omega \hat{\bfk} = \omega \hat{\bfx} = \omega \bfx/|\bfx|$, the transverse-traceless projector is defined as
\begin{align}
\Lambda_{ij,lm}(\hat{k})
&=  \delta_{il}\delta_{jm}-2 \hat k_j \hat k_m\delta_{il}+\frac{1}{2}\hat k_i \hat k_j \hat k_l \hat k_m -\frac{1}{2} \delta_{ij} \delta_{lm}+\frac{1}{2} \delta_{ij}\hat k_l \hat k_m +\frac{1}{2}\delta_{lm}\hat k_i \hat k_j
\end{align}
and the Fourier transform of the energy momentum tensor is
\be \label{Tfouriertr}
T_{ij}(\omega,\bfk) =  \int \ud^4 x e^{i\omega t-i\bfk\cdot \bfx}  T_{ij}(t,\bfx)  .
\ee
Since $\Lambda_{ij,lm}(\hat{k})$ projects out the trace part of $T_{ij}(\omega,\bfk)$, we can drop the manifest trace term $\delta_{ij} \mathcal{L}$ in $T_{ij}(t,\bfx)$ to simplify the calculation. Note that this formula has taken into account the conservation of the energy-momentum tensor and thus the scalar's equation of motion, as for a scalar its equation of motion is equivalent to the conservation of its energy-momentum tensor.

If there are only some isolated frequencies $\omega_N(N=1,2,...)$ in the energy-momentum tensor
\be
T_{\mu\nu}(\omega,\bfk) \equiv  \sum_N \tilde{T}_{\mu\nu}(\omega,\bfk) 2\pi\delta(\omega-\omega_N)  ,
\ee
then we can rewrite Eq.~(\ref{formula1}) as
\be  \label{formula3}
\frac{\ud P}{\ud \Omega} =  \sum_N \frac{G \omega_N^2}{\pi} \Lambda_{ij,lm}(\hat{k}_N) \tilde{T}^{ij}(\omega_N,\bfk_N) (\tilde{T}^{lm}(\omega_N,\bfk_N))^*
\ee
where again $\hat{\bfk}_N = \omega_N \hat{\bfx}$ and $P= E / {\rm time}$ is the power of the energy emission. Because of the projection $\Lambda_{ij,lm}(\hat{k})$, once the field profile is known, the potential of the scalar field, which only appears as a trace in $T_{ij}$, is not needed in the estimate of the gravitational wave production. The potential information in a sense is already encoded in the field profile.  

\section{Deviation from spherical symmetry} \label{sec:nssl}

Oscillons are most likely not perfectly spherical, especially at formation. Here we show that the suppression is still significant if the deviation from the spherical symmetry is not dramatic.

\begin{figure}
\begin{center}
\includegraphics[height=2in,width=4in]{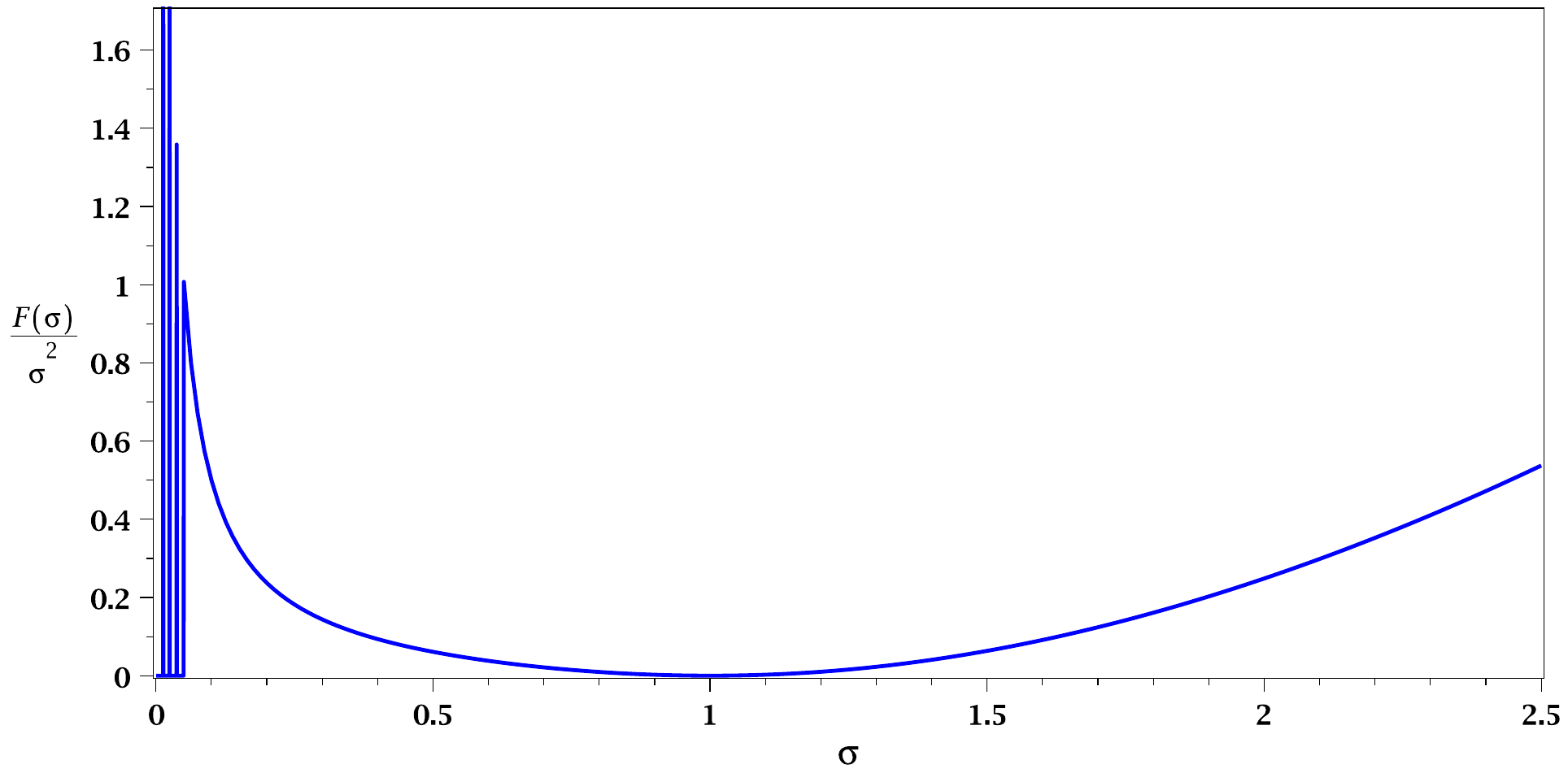}
\caption{The function $F(\sigma)/\sigma^2$. $m_x = m_y =\mu, m_z=\sigma \mu$, where $\sigma$ encodes the extent of the ellipse of the oscillating oscillon. For $\sigma=1$, we have $F(1)=0$, which means the oscillon is spherically symmetric and around which the gravitational energy emission is suppressed.} \label{figfalpha}
\end{center}
\end{figure}

We can imagine an infinite number of oscillons homogeneously distributed across the space, with the two-oscillon and four-oscillon contributions being exponentially suppressed, as with the spherically symmetric case. So, here we focus on the contribution from a single oscillon and adopt the following ansatz
\be \label{ansatz2}
\phi(t,{\bf x}) = A e^{ -\frac12 (m_x^2 x^2+m_y^2 y^2+m_z^2 z^2) }\cos( \mu t) ,
\ee
where $m_x$, $m_y$ and $m_z$ are in general different, thereby breaking spherical symmetry. Again, after some Fourier transforms and algebra, we obtain the power per solid angle as
\begin{align} \label{pnss}
\frac{\ud P_{2\mu}}{\ud \Omega} & =  \frac{2\pi^2 G A^4\mu^2}{(m_x m_y m_z)^2} \mathcal{F}(m_i,\hat{x}_i)    ,
\end{align}
where
\be
\mathcal{F}(m_i,\hat{x}_i) = \left[\left(\sum_i m_i^4(1-2\hat{x}_i^2)\right) + \left(\sum_i m_i^2 \hat{x}_i^2\right)^2 -\frac12 \left(\sum_i m_i^2(\hat{x}_i^2-1)\right)^2 \right] e^{-4\mu^2\sum_i\frac{\hat{x}_i^2}{m_i^2}}   .
\ee
The integration $\int \ud \Omega \mathcal{F}(m_i,\hat{x}_i)$ is difficult to do analytically for general $m_x$, $m_y$, $m_z$ and $\mu$. As a representative example, we can consider the case that the oscillon is elliptical, that is,
\be
m_x=m_y=\mu, ~~~~~~m_z=\sigma \mu   ,
\ee
where the dimensionless $\sigma$ encodes the extent of the ellipse. For this case, we have
\be
\int \ud \Omega \mathcal{F}(m_i,\hat{x}_i) = \mu^4 F(\sigma)  ,
\ee
with
\be
F(\sigma) = \frac{\pi}{32}e^{-4/\sigma^2}\sigma^2(-11\sigma^2 + 8) + \frac{\pi^{3/2}}{64e^4}
(83\sigma^4-144\sigma^2+64)\frac{{\rm erf} \left(\sqrt{4/\sigma^2-4}\right)}{\sqrt{4/\sigma^2-4}}   ,
\ee
where ${\rm erf}()$ is the error function. The integrated power of the gravitational wave emission for an elliptical oscillon is given
\begin{align} \label{powernonss}
P_{2\mu}  = 2\pi^2 G A^4 \frac{ F(\sigma)}{\sigma^2}   ,
\end{align}
where $F(\sigma)/\sigma^2$ is numerically given by Fig.~(\ref{figfalpha}). We can see that the curve is very flat around $\sigma=1$, the spherically symmetric case. So  suppression in the gravitational wave production is still significant even if the oscillon shape slightly deviates from the case of spherical symmetry.


\end{document}